\documentclass[12pt,a4paper]{article}

\usepackage{amssymb, latexsym, amsthm}
\usepackage{graphicx} 
\usepackage{url}      
\newcommand{\doi}[1]{\url{http://dx.doi.org/#1}}
\usepackage{amsmath}  

\IfFileExists{ajr.sty}{\usepackage{ajr}}{}
\usepackage{parskip}

\renewcommand{\phi}{\varphi}

\newcommand{\R}{{\mathbb R}}

\newcommand{\Ord}[1]{{\cal O}\big(#1\big)}

\newtheorem{theorem}{Theorem}[section]
\newtheorem{lemma}[theorem]{Lemma}
\newtheorem{coro}[theorem]{Corollary}
\newtheorem{defn}[theorem]{Definition}
\newenvironment{remark}{\paragraph{Remark}}{\par}

\newcommand{\given}{\operatorname{|}}
\newcommand{\Given}{\operatorname{\Big|}}

\title{Slow manifold and averaging for slow-fast stochastic differential system}

\author{
W. Wang\thanks{School of Mathematics, University of Adelaide,
South Australia, \textsc{Australia}. \protect\url{mailto:
w.wang@adelaide.edu.au}; and Department of
Mathematics, Nanjing University, Nanjing, \textsc{China}.
\protect\url{mailto:wangweinju@yahoo.com.cn}}
\and
A.~J. Roberts\thanks{School of Mathematics, University of Adelaide,
South Australia, \textsc{Australia}. \protect\url{mailto:
anthony.roberts@adelaide.edu.au}}
}

\date{\today}

\begin{document}

\maketitle

\begin{abstract}
We consider multiscale stochastic dynamical systems.
In this article an \emph{intermediate} reduced model
is obtained for a slow-fast system with fast mode driven
by white noise. First, the reduced stochastic system on
exponentially attracting slow manifold reduced system is
derived to errors of~$\mathcal{O}(\epsilon)$.
Second, averaging derives an autonomous deterministic
system up to errors of~$\mathcal{O}(\sqrt{\epsilon})$.
 Then an intermediate reduced model, which is an autonomous
 deterministic system driven by white noise up to errors
 of~$\mathcal{O}(\epsilon)$, is derived  using a martingale
 approach to account for fluctuations about the averaged
 system. This intermediate reduced model has a simpler
 form than the reduced model on the stochastic slow manifold.  These results not only connect averaging with the slow manifold, they also provide a martingale method for improving averaged models of stochastic systems.
\end{abstract}


\vskip 0.2cm \hspace{1cm}{\bf MSC}:\, 34C15; 37H10; 60H10

\section{Introduction}\label{sec:intro}
Complex dynamical systems in science and engineering often have multiple, disparate, time scales.  Here we consider the common case of two
widely separated time scales.  Recall that finite dimensional, deterministic systems are often written as the following slow-fast system of differential equations
\begin{eqnarray}
\dot{x}^\epsilon&=&Ax^\epsilon+f(x^\epsilon, y^\epsilon),\label{e:x}\\
\dot{y}^\epsilon&=&\frac{1}{\epsilon}B y^\epsilon+\frac{1}{\epsilon}g(x^\epsilon, y^\epsilon),\label{e:y}
\end{eqnarray}
where $\epsilon>0$ is a small parameter, $x^\epsilon(t)\in\R^n$ and $y^\epsilon(t)\in\R^m$ are the state variables, $A:\R^n\rightarrow\R^n$ and
$B:\R^m\rightarrow\R^m$ are bounded linear operators, and
$f:\R^n\times \R^m\rightarrow \R^n$ and
$g:\R^n\times\R^m\rightarrow \R^m$ are sufficiently smooth.
For small $\epsilon$, the averaging
principle suggests that a good approximation of the slow motion of the slow variables~$x^\epsilon$ on long time intervals is obtained by averaging~\eqref{e:x} over
the distribution of the fast variables~$y^\epsilon$ \cite[e.g.]{Arn, Bog61, Vol62}. The theory of invariant manifolds also supports, after the decay of exponential transients, the reduction of the dynamics of the system~(\ref{e:x})--(\ref{e:y}) to the dynamics of just~$x^\epsilon(t)$ \cite[e.g.]{Fen79, SS88} which being lower dimensional is simpler to represent and analyse.

For such deterministic systems, averaging is close to the slow manifold dynamics.
To derive a simplified system, the
invariant manifold method seeks to construct a lower dimensional, smooth,
attracting invariant set which gives a lower dimension reduced system by restricting
the system to this manifold. If the gap in the spectra of $A$~and~$B/\epsilon$
is large enough, then an exponentially attractive slow manifold
in the form $y=h(x)+\mathcal{O}(\epsilon)$ may be proved provided $f$~and~$g$ are Lipschitz. Then, to errors typically of~$\mathcal{O}(\epsilon)$, the lower dimensional, reduced system is
\begin{equation}\label{e:x-Mani-reduced}
\dot{x}=Ax+f(x, h(x)).
\end{equation}
On the other hand, in situations when the above slow manifold exists, the graph $y=h(x)$ is the unique fixed point,  exponentially stable, of the fast dynamics~(\ref{e:y})
with frozen $x\in\R^n$. Then, by the averaging principle, an
averaged system exists and indeed  is the reduced system~(\ref{e:x-Mani-reduced})~\cite{GKS04}. That is, in this deterministic case, if the system is autonomous,
the leading approximation to the slow manifold reduced system coincides with the
averaged system.

However, in the stochastic dynamics explored in this article, the above close coincidence no longer holds.
We explore the case when the fast variables~$y^\epsilon$ are perturbed by a stochastic force modelled by white noise~$\dot{W}$; that is, replace~\eqref{e:y} by
\begin{equation}\label{e:ye}
\dot{y}^\epsilon=\frac{1}{\epsilon}B y^\epsilon+\frac{1}{\epsilon}g(x^\epsilon, y^\epsilon)
+\frac{\sigma}{\sqrt{\epsilon}}\dot{W}\,.
\end{equation}
In this case the previously used fixed point of~(\ref{e:y}) instead becomes~$h^\epsilon(x, t)$,
an ergodic stationary process of the stochastic dynamics~(\ref{e:ye})
for frozen $x\in\R^n$.
Section~\ref{sec:reduction on IM} shows
that averaging obtains, by replacing~$h(x)$ with~$h^\epsilon(x, t)$ in~(\ref{e:x-Mani-reduced}), a random slow manifold reduced
model. In this case
the reduced system is a nonautonomous system with random
\emph{stationary} coefficient.
Thus,  the averaged system is a deterministic
autonomous system with $f(x, h(x)))$ in~(\ref{e:x-Mani-reduced})
replaced by
\begin{equation*}
\bar{f}(x)=\lim_{T\rightarrow\infty}\frac{1}{T}\int_0^Tf(x, h^\epsilon(x,t))\,dt\,.
\end{equation*}
The ergodic property of~$h^\epsilon(x, t)$, Section~\ref{sec:average}, ensures that the effective nonlinearity~$\bar{f}$ is well defined. Since a random slow manifold, reduced system generically contains stochastic effects,  necessarily the averaged system is different. The two approximations contain
different information of the stochastic dynamics of the original system.

Analysing a stochastic system is considerably more difficult than a
deterministic autonomous system. Thus the main goal of this article
is to derive a reduced model intermediate between the simple but
deficient averaged model and the complexity of the random slow
manifold model. The intermediate reduced model has
errors~$\Ord{\epsilon}$ when compared to original system, whereas
the averaged model has errors~$\Ord{\epsilon^{1/2}}$, see section
\ref{sec:average}. A key step here is to estimate  the fluctuation
of the slow manifold reduced system from the averaged system, which
is proved to be a Gaussian process by a martingale approach.


Section~\ref{sec:reduction
on IM} gives only the first order asymptotic approximation in~$\epsilon$ to the
random invariant manifold. For an example, Section~\ref{sec:toy model} illustrates the results via a simple example model.
Higher order terms in such an asymptotic approximation can be obtained by a normal form coordinate transformation~\cite{Rob08}.

Schmalfu\ss\ and Schneider~\cite{Sch08} recently studied the slow manifold of a
slow-fast random dynamical systems by a random graph transformation
on a slow time scale~$\mathcal{O}(\epsilon)$.  The random slow
manifold is shown to be a family of random fixed points for the fast
system~(\ref{e:ye}) parameterized by~$x$  with $\epsilon=1$\,. But this approach does
not give much information of the behavior of the slow system on the long time scales we aim for here.


\section{Preliminaries}
Consider the following stochastic slow-fast system:
\begin{eqnarray}\label{e:s-x}
\dot{x}^\epsilon&=&Ax^\epsilon+f(x^\epsilon, y^\epsilon),
\quad x^\epsilon(0)=x_0\in\R^n , \\
\dot{y}^\epsilon&=&\frac{1}{\epsilon}By^\epsilon+\frac{1}{\epsilon}g(x^\epsilon,
y^\epsilon)+\frac{\sigma}{\sqrt{\epsilon}}\dot{W}\,,
\quad y^\epsilon(0)=y_0\in\R^m , \label{e:s-y}
\end{eqnarray}
with matrix $A: \R^n\rightarrow \R^n$\,, matrix $B: \R^m\rightarrow \R^m$,
and nonlinear interaction functions $f: \R^n\times \R^m\rightarrow \R^n$, $g: \R^n\times
\R^m\rightarrow \R^m$ are $C^k$-smooth, $k\geq 2$\,, with
$f(0, 0)=f_x(0, y)=0$\,, $g(0, 0)=g_y(x, 0)=0$\,.  Here $\sigma$~is
a nonzero constant and $1\geq\epsilon>0$ is a parameter. $\{W(t),\,
t\in\R\}$ is a two-sided $\R^m$-valued Wiener process. Moreover, $f$~and~$g$ are Lipschitz with Lipschitz constant~$L_f$ and~$L_g$
respectively. Denote by~$|\cdot|_{\R^n}$\,, $|\cdot|_{\R^m}$  and~$|\cdot|_{\R^n}+|\cdot|_{\R^m}$ the distance defined on~$\R^n$,
$\R^m$ and~$\R^{n+m}$ respectively. For matrices~$A$ and~$B$ we assume
 \begin{description}
  \item[$\text{H}_1$] there are constants $\alpha\geq 0>\beta$ such that for any $x\in\R^n$ and $y\in\R^m$
\begin{equation}
|e^{tA}x|_{\R^n}\leq e^{\alpha t}|x|_{\R^n}\,, \quad t\leq 0\,,\quad
\text{and} \quad |e^{tB}y|_{\R^m}\leq e^{\beta t}|y|_{\R^m}\,, \quad
t\geq 0\,;
\end{equation}
 \item[$\text{H}_2$] the following spectral gap condition holds for any $1\geq\epsilon>0$\,,
\begin{equation}
L_f\frac{1}{\alpha-\lambda}+L_g\frac{1}{\epsilon\lambda-\beta}<1 \,,
\end{equation}
for some~$\lambda$ with $\beta<\epsilon\lambda<\alpha$\,.
 \end{description}

Under the above assumptions we give some basic results on
($x^\epsilon(t), y^\epsilon(t)$), the solutions of the slow-fast system~(\ref{e:s-x})--(\ref{e:s-y}).
First we have the following bounded estimation.
\begin{lemma}\label{lem:prior-est}
For any $(x_0, y_0)\in\R^{n+m}$, there is a unique solution
$(x^\epsilon(t),\, y^\epsilon(t))$ to the slow-fast system~(\ref{e:s-x})--(\ref{e:s-y}) such that
for any $T>0$
\begin{equation*}
(x^\epsilon,\, y^\epsilon)\in L^2(\Omega, C(0, T; \R^{n+m}))\,.
\end{equation*}
Moreover there is a positive constant~$C_T$ such that
\begin{equation*}
\mathbb{E}\sup_{0\leq t\leq T}|(x^\epsilon(t), y^\epsilon(t))|^2_{\R^{n+m}}
\leq C_T|(x_0, y_0)|^2_{\R^{n+m}}\,.
\end{equation*}
\end{lemma}
\begin{proof}
The existence and uniqueness result follows from a classical
approach~\cite{Arn0}.
By the assumption on~$f$ we have
\begin{equation*}
 \frac{1}{2}\frac{d}{dt}|x^\epsilon(t)|^2_{\R^n}\leq
 \|A\|_{\R^n}|x^\epsilon(t)|^2_{\R^n}+L_f|x^\epsilon(t)|^2_{\R^n}\,,
\end{equation*}
where $\|A\|_{\R^n}$ is the norm of~$A$ as a linear operator on~$\R^n$.
Also by the assumption on~$g$ and~$\textbf{H}_2$ we have
by It\^o's formula
\begin{equation*}
 \frac{1}{2}\frac{d}{dt}|y^\epsilon(t)|^2_{\R^m}\leq
 \frac{1}{\epsilon}\beta|y^\epsilon(t)|^2_{\R^m}+\frac{1}{\epsilon}L_g|y^\epsilon(t)|^2_{\R^m}
+\frac{\sigma^2}{2\epsilon}+\frac{1}{\epsilon}\langle y^\epsilon(t), \sigma \dot{W}\rangle_{\R^n}\,.
\end{equation*}
Then by the Burkholder--Davis--Gundy inequality~\cite{Huang} we have
the bounded estimation result.  This completes the proof.
\end{proof}
Now denote by~$\mu^\epsilon$ the probability measure defined by the distribution
 of the slow modes~$x^\epsilon$ on space~$C(0, T; \R^n)$\,. Then we have
the following tightness result.
\begin{theorem}\label{thm:tight}
The family of probability measure~$\{\mu^\epsilon\}_\epsilon$ is tight in space
$C(0, T; \R^n)$ for any $T>0$\,.
\end{theorem}
\begin{proof}
By~(\ref{e:s-x})
\begin{equation*}
 x^\epsilon(t)=x_0+\int_0^tAx^\epsilon(s)\,ds+\int_0^tf(x^\epsilon(s), y^\epsilon(s))\,ds\,.
\end{equation*}
For any $0<s<t$ we have
\begin{equation*}
\mathbb{E}\left|\int_s^tAx^\epsilon(s)\,ds\right|_{\R^n}\leq
\sqrt{t-s}\left[\int_s^t\mathbb{E}|Ax^\epsilon(s)|^2_{\R^n}\,ds\right]^{1/2}
\end{equation*}
and
\begin{equation*}
\mathbb{E}\left|\int_s^tf(x^\epsilon(s), y^\epsilon(s))\,ds\right|_{\R^n}
\leq \sqrt{t-s}\left[\int_s^t\mathbb{E}|f(x^\epsilon(s), y^\epsilon(s))|^2_{\R^n}\,ds \right]\,.
\end{equation*}
Then by the estimates of Lemma~\ref{lem:prior-est},
 the assumptions on $A$~and~$f$, the family
$\{x^\epsilon\}_\epsilon\subset L^2(\Omega, C^{1/2}(0, T; \R^n))$ is uniformly bounded with
respect to~$\epsilon$, which yields the tightness of~$\{\mu^\epsilon\}_\epsilon$ in $C(0, T; \R^n)$. The proof is complete.
\end{proof}

\section{Random slow manifold reduction}\label{sec:reduction on IM}

In this section a random slow manifold reduced system is derived for the slow-fast system~(\ref{e:s-x})--(\ref{e:s-y}) for any small $\epsilon\geq0$\,. We apply the theory of random invariant manifolds to
study the asymptotic behavior of the system~(\ref{e:s-x})--(\ref{e:s-y}). Theorem~\ref{thm:reduction} proves that the long time
behavior of~(\ref{e:s-x})--(\ref{e:s-y}) is described by the flow on
a random invariant manifold which is exponentially stable. Then by an asymptotic approximation, a random slow manifold reduced system is constructed.

Recall some basic concepts in random dynamical systems
(\textsc{rds}s) and random invariant manifolds. We work on the
canonical probability space $(\Omega, \mathcal{F}, \mathbb{P})$ with
$\Omega$~consisting of the sample paths of~$W(t)$. To be more
precise, $W$~is the identity on~$\Omega$, with
\begin{equation*}
\Omega=\{w\in C([0, \infty), \R^m): w(0)=0 \},
\end{equation*}
and $\mathbb{P}$ the Wiener measure~\cite{Arn98}.

Let $\theta_t: (\Omega, \mathcal{F},
\mathbb{P})\rightarrow (\Omega, \mathcal{F}, \mathbb{P}) $ be a
metric dynamical system (driven system), that is,
\begin{itemize}
    \item $\theta_0=\text{id}$,
    \item $\theta_t\theta_s=\theta_{t+s}$ for all~$s$,
        $t\in\mathbb{R}$\,,
    \item the map $(t,\omega)\mapsto \theta_t\omega$ is
     measurable and $\theta_t\mathbb{P}=\mathbb{P}$ for all
        $t\in\mathbb{R}$\,.
\end{itemize}
On $\Omega$ the map~$\theta_t$ is  the shift
\begin{equation}\label{MD}
\theta_t\omega(\cdot)=\omega(\cdot+t)-\omega(t),\quad t\in
\mathbb{R}\,, \quad\omega\in\Omega\,.
\end{equation}

\begin{defn}
\label{def:RDS} Let $(\mathcal{X}, d)$ be a metric space with Borel
$\sigma$-algebra~$\mathcal{B}$, then a random dynamical
system (\textsc{rds}) on~$(\mathcal{X}, d)$ over~$\theta_t$ on~$(\Omega, \mathcal{F}, \mathbb{P})$ is a measurable map
\begin{equation*}
\varphi:\mathbb{R}^+\times \Omega\times \mathcal{X}\rightarrow
\mathcal{X} \,,\quad
(t, \omega, x) \mapsto \varphi(t, \omega)x
 \end{equation*}
having the following cocycle property
\begin{equation*}
\varphi(0, \omega) x=x\,,\quad \varphi(t,
\theta_\tau\omega)\circ\varphi(\tau, \omega) x=\varphi(t+\tau,
\omega)x
 \end{equation*}
 for $t, \tau\in\mathbb{R}^+$, $x\in \mathcal{X}$ and $\omega\in\Omega$\,.
\end{defn}

A \textsc{rds}~$\varphi$ is continuous or differentiable if
$\varphi(t, \omega) : \mathcal{X}\rightarrow \mathcal{X}$ is
continuous or differentiable, respectively~\cite{Arn98}.

For a continuous random dynamical system~$\varphi$ on~$\mathcal{X}$
given by Definition~\ref{def:RDS}, we need the following  notions to
describe its dynamical behavior~\cite{CF94}.

\begin{defn}
A collection $M=M(\omega)_{\omega\in\Omega}$ of non-empty closed
sets $M(\omega)\subset \mathcal{X}$, $\omega\in\Omega$\,,  is a random  set
if
\begin{equation*}
\omega\mapsto\inf_{y\in M(\omega)}d(x, y)
\end{equation*}
is a real valued random variable for any $x\in \mathcal{X}$\,.
\end{defn}

\begin{defn}
A random set~${\cal B}(\omega)$ is called a tempered absorbing set for a
random dynamical system~$\varphi$ if for any bounded set~$K\subset\mathcal{X}$ there exists~$t_K(\omega)$ such that for all $t\geq t_K(\omega)$
\begin{equation*}
\varphi\big(t,\theta_{-t}\omega,K\big)\subset {\cal B}(\omega),
\end{equation*}
and for all $\varepsilon>0$
\begin{eqnarray*}
 \lim_{t\rightarrow\infty}e^{-\varepsilon
t}d\big({\cal B}(\theta_{-t}\omega)\big)=0\,, \quad \text{for a.e. }  \omega\in \Omega\,,
\end{eqnarray*}
where $d({\cal B})=\sup_{x\in {\cal B}}d(x, 0)$, with $0\in\mathcal{X}$, is the
diameter of~${\cal B}$, if $0\in {\cal B}$.
\end{defn}
Then we introduce the random invariant manifold~\cite{Arn98,
DLSch04}.
\begin{defn}
A random set~$M(\omega)$ is called a positive invariant set for a
random dynamical system~$\varphi(t,\omega,x)$ if
\begin{equation*}
\varphi(t,\omega, M(\omega))\subset M(\theta_t\omega),\quad\text{for }t\geq
0\,.
\end{equation*}
If $M(\omega)=\{x_1+\psi(x_1,\omega):x_1\in \mathcal{X}_1 \}$
is the  graph of a random Lipschitz mapping
\begin{equation*}
\psi(\cdot,\omega): \mathcal{X}_1\rightarrow \mathcal{X}_2
\end{equation*}
with $\mathcal{X}=\mathcal{X}_1\oplus \mathcal{X}_2$,
then $M(\omega)$ is called a Lipschitz invariant manifold of
$\varphi$.
\end{defn}
Duan et al.~\cite{DLSch04} give further details about random invariant manifold theory.

For our purpose we introduce a stationary process~$\eta^\epsilon(t)$
which solves the linear stochastic differential equation
\begin{equation}\label{e:OU}
d\eta^\epsilon(t)=\frac{1}{\epsilon}B\eta^\epsilon(t)\,
dt+\frac{\sigma}{\sqrt{\epsilon}}\,dW(t)\,.
\end{equation}
Then by the assumption~$(\textbf{H}_1)$, $\eta^\epsilon(t)$~is
exponential mixing. Moreover, under the driven system we write the
stationary  process as~$\eta^\epsilon(\theta_t\omega)$. Denote by
$\theta^\epsilon_t=\theta_{{t}/{\epsilon}}$ the scaled version
metric dynamical system $\theta_t$\,. Then, by Lemma~3.2 proved
by Schmalfu\ss\  et al.~\cite{Sch08},
$\eta^\epsilon(\theta_t\omega)$ has the same distribution of~$\eta(\theta^\epsilon_t\omega)$ with $\eta(t)=\eta(\theta_t\omega)$
solving the following linear stochastic differential equation
\begin{equation}\label{e:OU1}
d\eta(t)=B\eta(t)\, dt+\sigma\,dW(t)\,.
\end{equation}




Now we consider the following slow-fast random differential system
\begin{eqnarray}
\dot{X}^\epsilon&=&AX^\epsilon+F(X^\epsilon, Y^\epsilon, \,\theta_t^\epsilon\omega), \quad X^\epsilon(0)=X_0\,,\label{e:r-X}\\
\dot{Y}^\epsilon&=&\frac{1}{\epsilon}BY^\epsilon+\frac{1}{\epsilon}G(X^\epsilon, Y^\epsilon,\,
\theta_t^\epsilon\omega)\,,\quad Y^\epsilon(0)=Y_0\,,\label{e:r-Y}
\end{eqnarray}
with
\begin{eqnarray}
F(X^\epsilon(t), Y^\epsilon(t), \theta_t^\epsilon\omega)&=&f(X^\epsilon(t),
Y^\epsilon(t)+\eta^\epsilon(\theta_t\omega)),\label{e:F}\\
G(X^\epsilon(t), Y^\epsilon(t), \theta_t^\epsilon\omega)&=&g(X^\epsilon(t),
Y^\epsilon(t)+\eta^\epsilon(\theta_t\omega)).\label{e:G}
\end{eqnarray}
By the assumption on~$f$ and~$g$,  solutions~$(X^\epsilon(t), Y^\epsilon(t))$ to~(\ref{e:r-X})--(\ref{e:r-Y}) define a continuous random
dynamical system~$\Phi^\epsilon(t, \omega)$ on~$\R^{n+m}$ defined by
\begin{equation*}
\Phi^\epsilon(t,\omega)(X_0, Y_0)=(X^\epsilon(t, \omega), Y^\epsilon(t, \omega)).
\end{equation*}
Then $\phi^\epsilon(t, \omega)=\Phi^\epsilon(t, \omega)+(0,
\eta^\epsilon(\theta_t\omega))$ is a continuous random dynamical
system defined by the slow-fast system~(\ref{e:s-x})--(\ref{e:s-y}).

Next we show that there is a random invariant set~$\mathcal{M}(\omega)$ for~$\Phi^\epsilon(t,\omega)$. We follow the
method of Lyapunov--Perron for random dynamical
systems~\cite{DLSch04}. For some
$\beta<\lambda<\alpha$ define the Banach space
\begin{equation*}
C^-_\lambda=\{u:(-\infty, 0]\rightarrow \R^{n+m}: u \text{ is
continuous and } \sup_{t\leq0}e^{-\lambda
t}|u(t)|_{\R^{n+m}}<\infty\}
\end{equation*}
with the norm
\begin{equation*}
|u|_{C^-_\lambda}=\sup_{t\in(-\infty, 0]}e^{-\lambda
t}|u(t)|_{\R^{n+m}}\,.
\end{equation*}
For any given $X_0\in\R^n$\,, define the map $\mathcal{T}(X_0, \omega,
\cdot)$ on~$ C^-_\lambda$ by
\begin{eqnarray}
\mathcal{T}\big(X_0, \omega, (X^\epsilon,
Y^\epsilon)\big)(t)&=&\left(e^{tA}X_0+\int^0_te^{(t-s)A}F(X^\epsilon(s), Y^\epsilon(s),
\theta_s^\epsilon\omega)\,ds\,,
\right.\nonumber\\&&\left.{}
\frac{1}{\epsilon}\int^t_{-\infty}e^{({t-s})B/{\epsilon}}G(X^\epsilon(s),
Y^\epsilon(s), \theta^\epsilon_s\omega)\,ds\right).\label{e:T}
\end{eqnarray}
Then, for any $X_0\in\R^n$\,, $\mathcal{T}(X_0, \omega, \cdot):
C_\lambda^-\rightarrow C_\lambda^-$. Moreover, for each $(X^\epsilon,
Y^\epsilon), (\tilde{X}^\epsilon, \tilde{Y}^\epsilon)\in
C^-_\lambda$ we have
\begin{eqnarray*}
&&|\mathcal{T}\big(X_0,\, \omega, (X^\epsilon,
Y^\epsilon)\big)-\mathcal{T}\big(X_0,\, \omega, (\tilde{X}^\epsilon,
\tilde{Y}^\epsilon)\big)|_{C^-_\lambda}\\&\leq& \sup_{t\leq
0}\Big|\int_0^te^{-\lambda t}e^{(t-s)A}\big[F(X^\epsilon(s), Y^\epsilon(s),
\theta_s^\epsilon\omega)-F(\tilde{X}^\epsilon(s), \tilde{Y}^\epsilon(s),
\theta_s^\epsilon\omega)\big] \Big|_{\R^n}\\&&{}+ \sup_{t\leq 0}\Big|\frac{1}{\epsilon}\int_{-\infty}^te^{-\lambda
t}e^{({t-s})B/{\epsilon}}\big[G(X^\epsilon(s), Y^\epsilon(s),
\theta_s^\epsilon\omega)-G(\tilde{X}^\epsilon(s), \tilde{Y}^\epsilon(s),
\theta_s^\epsilon\omega)\big] \Big|_{\R^m}\\
&\leq&\left[L_f\int^0_te^{(-\lambda+\alpha)(t-s)}\,ds+\frac{L_g}{\epsilon}
\int_{-\infty}^te^{({\beta-\epsilon\lambda})(t-s)/{\epsilon}}\,ds\right]\big|(X^\epsilon,
Y^\epsilon)-(\tilde{X}^\epsilon,
\tilde{Y}^\epsilon)\big|_{C^-_\lambda} \\
&\leq&\left[L_f\frac{1}{\alpha-\lambda}+L_g\frac{1}{\epsilon\lambda-\beta}\right]
\big|(X^\epsilon, Y^\epsilon)-(\tilde{X}^\epsilon, \tilde{Y}^\epsilon)\big|_{C^-_\lambda}\,.
\end{eqnarray*}
Then by Assumption~$\textbf{H}_2$, and the Banach fixed point
theorem, for any $X_0\in\R^n$, $\mathcal{T}^\epsilon$ has a unique fixed
point $(X^{\epsilon*}, Y^{\epsilon*})\in C^-_\lambda$ with
\begin{eqnarray*}
X^{\epsilon*}(t)&=&e^{tA}X_0+\int^0_te^{(t-s)A}F(X^{\epsilon*}(s), Y^{\epsilon*}(s),
\theta_s^\epsilon\omega)\,ds\,,\\
Y^{\epsilon*}(t)&=&\frac{1}{\epsilon}\int^t_{-\infty}e^{({t-s})B/{\epsilon}}G(X^{\epsilon*}(s),
Y^{\epsilon*}(s), \theta^\epsilon_s\omega)\,ds\,.
\end{eqnarray*}
Now for each $\omega\in\Omega$\,, define a map $h^\epsilon(\cdot,
\omega):\R^n\rightarrow\R^m$ by
\begin{equation}\label{e:h}
X_0\mapsto h^\epsilon(X_0,
\omega)=\frac{1}{\epsilon}\int_{-\infty}^0e^{-B s/\epsilon}G(X^{\epsilon*}(s),
Y^{\epsilon*}(s), \theta^\epsilon_s\omega)\,ds
\end{equation}
which is Lipschitz continuous with Lipschitz constant
\begin{equation*}
L_h=\frac{L_g}{(\alpha-\lambda)\big\{1-L_g\big[1/(\alpha-\lambda)+
1/(\epsilon\lambda-\beta)\big]\big\}}\,.
\end{equation*}
Then by a similar discussion to that of Duan et al.~\cite{DLSch04} we draw the
following result.
\begin{theorem}
Suppose assumptions~$\mathbf{H}_1$ and~$\mathbf{H}_2$. There exists
a Lipschitz continuous random invariant manifold~$\mathcal{M}^\epsilon(\omega)$ for~$\Phi^\epsilon(t,\omega)$ which is
$\mathcal{M}^\epsilon(\omega)=\{(X,\, h^\epsilon(X, \omega)):X\in\R^n\}$.
\end{theorem}
\begin{remark}
The random manifold~$\mathcal{M}^\epsilon(\omega)$ is independent of the
choice of~$\lambda$ which satisfies assumption~$\mathbf{H}_2$.
\end{remark}

By the definition of~$\phi^\epsilon(t, \omega)$ we have the following
corollary.
\begin{coro}
Suppose assumptions~$\mathbf{H}_1$ and~$\mathbf{H}_2$. The fast-slow system~(\ref{e:s-x})--(\ref{e:s-y}) has a Lipschitz continuous random
invariant manifold which is written as
$\mathcal{M}^\epsilon(\omega)+(0, \eta^\epsilon(\omega))$\,.
\end{coro}

\begin{remark}\label{rem:h-e}
 From~(\ref{e:h}), by the transformation
$s/\epsilon\rightarrow s$
\begin{eqnarray}\label{e:h-slow}
 h^\epsilon(X, \omega)&=&\int_{-\infty}^0e^{-sB}G(X^{\epsilon*}( \epsilon s),
Y^{\epsilon*}(\epsilon s), \theta^\epsilon_{\epsilon
s}\omega)\,ds\nonumber\\ \text{(in distribution)}
&=&\int_{-\infty}^0 e^{-sB}G(\bar{X}^{\epsilon*}(s),
\bar{Y}^{\epsilon*}(s), \theta_s\omega)\,ds
\end{eqnarray}
where
\begin{eqnarray*}
\bar{X}^{\epsilon*}(t)&=&e^{\epsilon
tA}X+\epsilon\int^0_te^{\epsilon(t-s)A}F(\bar{X}^{\epsilon*}(s), \bar{Y}^{\epsilon*}(s),
\theta_s\omega)\,ds\,,\quad t\leq 0\,,\\
\bar{Y}^{\epsilon*}(t)&=&\int^t_{-\infty}e^{(t-s)B}G(\bar{X}^{\epsilon*}(s),
\bar{Y}^{\epsilon*}(s), \theta_s\omega)\,ds\,,\quad t\leq 0
\end{eqnarray*}
is the unique solution in~$C^-_{\epsilon\lambda}$ of the following system with a slow time scale
\begin{eqnarray}
\dot{X}^\epsilon(t)&=&\epsilon AX^\epsilon(t)+\epsilon F(X^\epsilon(t), Y^\epsilon(t), \,\theta_t\omega),\label{e:slowr-X}\\
\dot{Y}^\epsilon(t)&=&BY^\epsilon(t)+G(X^\epsilon(t), Y^\epsilon(t),\,
\theta_t\omega).\label{e:slowr-Y}
\end{eqnarray}
This can be deduced by the same approach to construct~$h^\epsilon(X, \omega)$. Then~$\mathcal{M}^\epsilon(\omega)$ is also an invariant manifold for system~(\ref{e:slowr-X})--(\ref{e:slowr-Y}) driven by~$\theta_t$.
\end{remark}

Now for small $\epsilon>0$ we give an asymptotic approximation for~$h^\epsilon(X, \omega)$
for fixed $\omega\in\Omega$\,. For each $X\in\R^n$, consider
\begin{equation}\label{e:Y-0}
\dot{Y}(t)=BY(t)+G(X, Y(t), \theta_t\omega)\,.
\end{equation}
By the assumption on~$g$ and~$\textbf{H}_1$, (\ref{e:Y-0})~has a unique solution in~$C^-_{\epsilon\lambda}$ for any $\epsilon>0$,
denoted by $\bar{Y}(t)=h^0(X,\theta_t\omega)$\,, $t\leq 0$\,,
with
\begin{equation*}
h^0(X, \omega)=\int^0_{-\infty}e^{-sB}G\left(X, h^0(X, \theta_s\omega),
\theta_s\omega\right)\,ds\,.
\end{equation*}
The existence of~$h^0(X, \omega)$ can be deduced by exactly
the same approach to construct~$h^\epsilon(X, \omega)$.
Then we have the following approximation result.
\begin{lemma}\label{lem:he-h0}
Assume~$\mathbf{H}_1$ and~$\mathbf{H}_2$ hold. Then for
 almost all $\omega\in\Omega$\,,
\begin{equation*}
|h^\epsilon(X, \omega)-h^0(X, \omega)|_{\R^m}=\mathcal{O}(\epsilon)\quad
\text{as } \epsilon\rightarrow 0\,,
\end{equation*}
which is uniformly for~$X$ in any bounded subset of~$\R^n$.
\end{lemma}
\begin{proof}
By Remark~\ref{rem:h-e}, for any $X\in\R^n$\,, $t\leq 0$\,,
\begin{eqnarray*}
&&\left|\bar{X}^{\epsilon*}(t)-X\right|_{C^-_{\epsilon\lambda}}\\&\leq&
|e^{(\epsilon A-\epsilon\lambda)t}X-e^{-\epsilon\lambda t}X|_{\R^n}+
\epsilon L_f\left|(\bar{X}^{\epsilon*}, \bar{Y}^{\epsilon*})\right|_{C_\lambda^-}\int^0_te^{(-\epsilon\lambda+\epsilon A)(t-s)}\,ds\,.
\end{eqnarray*}
Then by the assumption~$\textbf{H}_1$ and~$\textbf{H}_2$,
\begin{equation*}
|\bar{X}^{\epsilon*}(t)-X|_{C^-_{\epsilon\lambda}}=\mathcal{O}(\epsilon)\quad
\text{as }\epsilon\rightarrow 0\,.
\end{equation*}
Further
\begin{eqnarray*}
&&|\bar{Y}^{\epsilon*}(t)-\bar{Y}(t)|_{C^-_{\epsilon\lambda}}\\
&\leq &\frac{L_g}{\epsilon\lambda-\beta}\left|\bar{X}^{\epsilon*}(t)-X\right|_{C^-_{\epsilon\lambda}}+
\frac{L_g}{\epsilon\lambda-\beta}\left|\bar{Y}^{\epsilon*}(t)-\bar{Y}(t))\right|_{C^-_{\epsilon\lambda}}\,.
 \end{eqnarray*}
Then  also  by assumption~$\textbf{H}_1$ and~$\textbf{H}_2$,
 for any $X\in\R^n$ and $\omega\in\Omega$\,,
\begin{equation*}
|h^\epsilon(X, \omega)-h^0(X, \omega)|_{\R^m} \leq
|\bar{Y}^{\epsilon*}(t)-\bar{Y}(t)|_{C^-_{\epsilon\lambda}}=\mathcal{O}(\epsilon),\quad
\epsilon\rightarrow 0\,.
\end{equation*}
This completes the proof.
\end{proof}
\begin{remark}
Similarly, by Remark~\ref{rem:h-e}, we have
\begin{eqnarray}
h^0(X,\omega)&=&\frac{1}{\epsilon}\int^0_{-\infty}e^{- B s/\epsilon}G(X, \bar{Y}^\epsilon(s),
\theta_{s/\epsilon}\omega)\,ds\nonumber\\
\text{(in distribution)}&=& \frac{1}{\epsilon}\int^0_{-\infty}e^{- B
s/\epsilon}G(X, \bar{Y}^\epsilon(s),
\theta^\epsilon_s\omega)\,ds\label{e:h-0}
\end{eqnarray}
where $(X, \bar{Y}^\epsilon(t))\in C^-_\lambda$ solves
\begin{equation}\label{e:Y-e}
\dot{Y}^\epsilon(t)=\frac{1}{\epsilon}BY^\epsilon(t)+\frac{1}{\epsilon}
G\left(X, Y^\epsilon(t), \theta^\epsilon_t\omega\right).
\end{equation}
Then $\mathcal{M}^0(\omega)=\{(X, h^0(X, \omega)):
X\in\R^n\}$ is an invariant manifold for systems~(\ref{e:Y-0}) and~(\ref{e:Y-e}) driven by~$\theta_t$ and~$\theta_t^\epsilon$ respectively.
\end{remark}

%

Given a random invariant manifold, in the following, if the spectral gap is large enough, any solution of~(\ref{e:r-X})--(\ref{e:r-Y})
is proved to converge exponentially quickly to a flow on the manifold~$\mathcal{M}^\epsilon(\omega)$ as $t\rightarrow \infty$\,. Then the
$(n+m)$-dimensional slow-fast system~(\ref{e:s-x})--(\ref{e:s-y}) is reduced
to an $n$-dimensional system. First we give the following definition
for a random dynamical system~$\Phi$.

\begin{defn} [Almost sure asymptotic completeness]\label{completeness}
 Let~$\mathcal{M}(\omega)$
be an invariant manifold for random dynamical system~$\Phi(t,\omega)$. The invariant manifold~$\mathcal{M}$ is called
almost surely asymptotically complete  if for every $ z\in
\R^{n+m}$, there exists $z'\in\mathcal{M}(\omega)$ such that
\begin{equation*}
|\Phi(t,\omega)z-\Phi(t,\omega)z'|\leq D(\omega)|z-z'|e^{-kt},
\quad t\geq 0\,,
\end{equation*}
for almost all $\omega\in\Omega$, where $k$~is some positive
constant and $D$~is a positive   random variable.
\end{defn}
Following the approach of Wang and Duan~\cite{WD07} we
have the following result.
\begin{theorem}\label{thm:reduction}
Assume~$\mathbf{H}_1$ and~$\mathbf{H}_2$. If there is $\delta>0$
such that
\begin{equation}\label{e:gap}
\epsilon\alpha+\epsilon L_f+\epsilon\delta L_f+\beta+L_g+\delta^{-1}L_g<0\,,
\end{equation}
then the random manifold~$\mathcal{M}(\omega)$ is almost surely
asymptotic complete. That is, for any solution~$(X^\epsilon(t, \omega),
Y^\epsilon(t, \omega)$ of~(\ref{e:r-X})--(\ref{e:r-Y}), there exists an
orbit $(\tilde{X}^\epsilon(t, \omega), \tilde{Y}^\epsilon(t, \omega))$ on the
manifold~$\mathcal{M}(\theta_t\omega)$ which is governed by the
following $n$-dimensional different equation
\begin{equation}\label{e:X-reduced}
\dot{\tilde{X}}^\epsilon=A\tilde{X}^\epsilon+F(\tilde{X}^\epsilon, \tilde{Y^\epsilon},
\theta^\epsilon_t\omega), \quad \tilde{Y}^\epsilon=h^\epsilon(\tilde{X}^\epsilon,
\theta_t\omega),
\end{equation}
such that for any $t\geq 0$ and almost sure $\omega\in\Omega$
\begin{eqnarray*}
&&|(X^\epsilon(t,\omega), Y^\epsilon(t, \omega))-(\tilde{X}^\epsilon(t, \omega),
\tilde{Y}^\epsilon(t, \omega))|_{\R^{n+m}}\\
&\leq& D|(X_0, Y_0)-(\tilde{X}^\epsilon(0),
\tilde{Y}^\epsilon(0))|_{\R^{n+m}}e^{-\gamma t/\epsilon}
\end{eqnarray*}
with $\gamma=-\beta-L_g-\delta^{-1}L_g>0$ and some deterministic
constant $D>0$\,.
\end{theorem}

\begin{remark}
An important problem is to predict the initial value~$(\tilde{X}^\epsilon(0), \tilde{Y}^\epsilon(0))$\,, which one can
deduce from a stochastic normal form~\cite{Rob08} or related dynamical considerations~\cite{Roberts89b}.
\end{remark}
By the assumption on~$f$, from the reduced system~(\ref{e:X-reduced}) and Lemma~\ref{lem:he-h0}, we have
\begin{eqnarray*}
\dot{\tilde{X}}^\epsilon&=&A\tilde{X}^\epsilon+F(\tilde{X}^\epsilon,
h^\epsilon(\tilde{X}^\epsilon, \theta_t^\epsilon\omega), \theta_t^\epsilon\omega)\\
&=&A\tilde{X}^\epsilon+F(\tilde{X}^\epsilon,
h^0(\tilde{X}^\epsilon, \theta_t^\epsilon\omega)+\mathcal{O}(\epsilon), \theta_t^\epsilon\omega)\\
&=& A\tilde{X}^\epsilon+F(\tilde{X}^\epsilon,
h^0(\tilde{X}^\epsilon, \theta_t^\epsilon\omega), \theta_t^\epsilon\omega)+\mathcal{O}(\epsilon).
\end{eqnarray*}
Then we draw the following slow manifold reduced result.
\begin{coro}
The slow manifold reduced system for the slow-fast system~(\ref{e:s-x})--(\ref{e:s-y}),
up to errors of~$\mathcal{O}(\epsilon)$, is
\begin{equation}\label{e:reduced}
\dot{x}^\epsilon(t)=Ax^\epsilon(t)+f\big(x^\epsilon(t), h^0(x^\epsilon(t),
\theta^\epsilon_t\omega)+\eta(\theta^\epsilon_t\omega)\big)\,.
\end{equation}
\end{coro}
\begin{remark}
On the other hand, by assumption~$\textbf{H}_2$, $L_g+\beta<0$
which yields that for any fixed $X\in\R^n$, system~(\ref{e:r-Y})
has a unique stationary solution which is,  by the definition,~$h^0(X, \omega)$. Moreover, the stationary solution is exponentially stable for almost
all $\omega\in\Omega$\,. So the slow manifold reduced system~(\ref{e:reduced}) is obtained by replacing~$y^\epsilon$ by
 $\bar{y}^{\epsilon, x}(\omega)=h^0(x,\omega)+\eta^\epsilon(\omega)$ the
stationary solution of~(\ref{e:s-y}) for fixed~$x$. This is the same as the deterministic autonomous case.
\end{remark}

\section{Application to a simple example model}\label{sec:toy model}
Consider the following nonlinear slow-fast stochastic system with $(x, y)\in\R^1\times\R^1$:
\begin{eqnarray}\label{e:toy-fast1}
dx&=&f(x, y)\,dt\,,\\ dy&=&\frac{1}{\epsilon}\left[-y+g(x,
y)\right]\,dt+\frac{\sigma}{\sqrt{\epsilon}}\,dW(t),\label{e:toy-fast2}
\end{eqnarray}
where~$f$ and~$g$ are smooth with the following specific values close to the origin, and far from the origin
\begin{equation*}
f(x, y)=  \begin{cases}
            -xy\,,&  x^2+y^2< 1/16\,,\\
             0\,,  &  x^2+y^2> 1/8\,,
          \end{cases}
\end{equation*}
and
\begin{equation*}
g(x, y)= \begin{cases}
            x^2-2y^2\,,&  x^2+y^2< 1/16\,,\\
             0\,,  & x^2+y^2> 1/8\,.
          \end{cases}
\end{equation*}
In a slow time scale we rewrite the above system as
\begin{eqnarray}\label{e:toy-slow1}
dx&=&\epsilon f(x, y)\,dt\\  dy&=&\left[-y+g(x,
y)\right]dt+\sigma\,dW(t). \label{e:toy-slow2}
\end{eqnarray}
The nonlinearities $f$~and~$g$ satisfy assumptions~$\textbf{H}_1$ and~$\textbf{H}_2$. Then by Theorem~\ref{thm:reduction}, the example system~(\ref{e:toy-fast1})--(\ref{e:toy-fast2}) has an invariant manifold
$\mathcal{M}(\omega)=\{(x, h^\epsilon(x,
\omega)+\eta^\epsilon(\omega))$\,, $x\in\R$\} and the reduced system
on this manifold is
\begin{equation}\label{e:toyreduced}
dx=f\left[x, h^\epsilon(x, \theta_t\omega)+\eta^\epsilon(\theta_t\omega)\right]dt\,.
\end{equation}
Further by the definition of $f$~and~$g$ and~(\ref{e:h-0}), $h^\epsilon$~has the following asymptotic expression near $x=0$
\begin{eqnarray*}
Y^{\epsilon*}(0)&=&h_0(x,
\omega)+\mathcal{O}(\epsilon)\\&=&\frac{1}{\epsilon}\int_{-\infty}^0
e^{ s/\epsilon}[x^2-2(h_0(x,
\theta^\epsilon_s\omega)+\eta(\theta^\epsilon_s\omega))^2]ds+\mathcal{O}(\epsilon)\\
&=&\frac{1}{\epsilon}\int_{-\infty}^0 e^{ s/\epsilon}
 [x^2-2(x^2+\eta(\theta^\epsilon_s\omega))^2]ds+\mathcal{O}(\epsilon, x^2)\\
&=&\frac{1}{\epsilon}\int_{-\infty}^0 e^{ s/\epsilon}[x^2-2(x^2+\eta(\theta^\epsilon_s\omega))^2]ds+\mathcal{O}(\epsilon, x^2)\\
&=&x^2-\frac{2}{\epsilon}\int_{-\infty}^0
e^{ s/\epsilon}[\eta(\theta^\epsilon_s\omega)]^2ds-
\frac{4x^2}{\epsilon}\int_{-\infty}^0
e^{ s/\epsilon}\eta(\theta^\epsilon_s\omega)ds-2x^4
\\&&{}+\mathcal{O}(\epsilon, x^3)
\end{eqnarray*}
Then an asymptotic expression of the reduced dynamics of the example system~(\ref{e:toyreduced}) is
\begin{align}\label{e:toyreduced0}
\dot{x}=&
-x^3-x\eta^\epsilon(\theta_t\omega)+\frac{2x}{\epsilon}\int_{-\infty}^0
e^{s/\epsilon}[\eta^\epsilon(\theta_{s+t}\omega)]^2ds+\frac{4x^3}{\epsilon}\int_{\infty}^0
e^{s/\epsilon}\eta^\epsilon(\theta_{t+s}\omega)ds
\nonumber\\&{}+\mathcal{O}(\epsilon, x^4).
\end{align}
For small~$\epsilon$ the random integrals in~(\ref{e:toyreduced0}) can be
replaced by some stochastic term on a long time scale as follows~\cite[e.g.]{Chao95}.
\begin{lemma}
For $\epsilon$ small, in a mean square sense
\begin{eqnarray*}
\frac{1}{\sqrt{\epsilon}}\eta^\epsilon(t)\rightarrow\sigma\,dW_1(t),\quad
\frac{1}{\epsilon\sqrt{\epsilon}}\int_{-\infty}^0e^{s/\epsilon}\eta^\epsilon(\theta_{s+t}\omega)ds
\rightarrow\sigma\,dW_2(t),
\end{eqnarray*}
and
\begin{equation*}
\frac{1}{\epsilon}\int_{-\infty}^0e^{s/\epsilon}[\eta^\epsilon(\theta_{s+t}\omega)]^2ds
\rightarrow \frac{\sigma^2}{2}+\sigma^2\sqrt{\frac{\epsilon }{2}}\,dW_3(t)
\end{equation*}
with some mutually independent standard scalar Brownian motions~$W_1(t)$, $W_2(t)$ and~$W_3(t)$.
\end{lemma}
\begin{proof}
This follows by a martingale approach. Let
\begin{equation*}
z^\epsilon(t)=\frac{1}{\sqrt{\epsilon}}\int_0^t\eta^\epsilon(s)\,ds\,,
\end{equation*}
then $z^\epsilon$ solves
\begin{equation*}
\dot{z}^\epsilon=\frac{1}{\sqrt{\epsilon}}\eta^\epsilon(t), \quad z^\epsilon(0)=0\,.
\end{equation*}
Then by  martingale results~\cite[e.g.]{Wa88, WR08}, we have~$z$, the limit of~$z^\epsilon$ as $\epsilon\rightarrow 0$, solves
\begin{equation*}
dz=\sigma \,dW_1\,, \quad z(0)=0\,,
\end{equation*}
for some scalar Brownian motion~$W_1(t)$. The others can be
obtained similarly.
\end{proof}

Thus for small~$\epsilon$,  the reduced system~(\ref{e:toyreduced}) is
\begin{align}\label{e:toyreduced1}
dx=& (-x^3+\sigma^2x)dt-\sqrt{\epsilon}\sigma
x\,dW_1(t)+\sqrt{2\epsilon}\sigma^2x\,
dW_3(t)+4\sqrt{\epsilon}\sigma x^3 dW_2(t)
\nonumber\\&{}+\mathcal{O}(\epsilon, x^4).
\end{align}
Observe that the three noise terms are all $1/2$-order in~$\epsilon$;
further, the three noise terms  can be replaced by one noise term.
This will be treated by averaging in next section.

Note that the reduced model~(\ref{e:toyreduced1}) is same as
that deduced by the stochastic normal form~\cite{Rob08} of the slow-fast system~\eqref{e:toy-fast1}--\eqref{e:toy-fast2}.

\section{Averaging approximation for small $\epsilon$}\label{sec:average}
The random invariant manifold reduction gives a lower dimensional system
which is a nonautonomous random system.
This section, for small~$\epsilon$,  derives a simpler reduced system which
is an autonomous system perturbed by a small stochastic term.
Moreover, the approximating errors prove to be of~$\mathcal{O}(\epsilon)$.
First, as $\epsilon$~is small in the slow-fast system~(\ref{e:s-x})--(\ref{e:s-y}),
$y^\epsilon$~is highly fluctuating
in time, and so an averaging method
derives the averaged equation~(\ref{e:averaged})
 which approximates~$x^\epsilon$ up to errors of~$\mathcal{O}(\sqrt{\epsilon})$. Second, by a martingale
approach, we derive an intermediate reduced equation~(\ref{e:inter-reduced})
which approximates~$x^\epsilon$ to errors of~$\mathcal{O}(\epsilon)$.

 We just need to consider the
reduced system~(\ref{e:reduced}). First define the following averaged equation
\begin{equation}\label{e:averaged}
\dot{x}(t)=Ax(t)+\bar{f}(x(t)), \quad x(0)=x_0\,,
\end{equation}
with the averaged
\begin{equation*}
\bar{f}(x)=\mathbb{E}f(x, \bar{y}^{\epsilon, x}(\omega))
\end{equation*}
which is independent of~$\epsilon$. By the Lipschitz property of~$h^\epsilon(x, \omega)$ and the assumption on~$f$, the average~$\bar f$ is Lipschitz; denote the Lipschitz
constant by~$L_{\bar{f}}$. Moreover, by the exponential mixing of~$\bar{y}^{\epsilon, x}$ we
have for any solution~$y^{\epsilon, x}(t)$ of~(\ref{e:s-y}) with initial
value $y_0\in\R^m$ and any fixed $x\in\R^n$
\begin{equation}\label{e:y-ybar}
\mathbb{E}|y^{\epsilon, x}(t)-\bar{y}^{\epsilon, x}(t)|^2_{\R^m}\leq
\mathbb{E}|y_0-\bar{y}^{\epsilon,x}(0)|^2e^{\beta t/\epsilon }\,.
\end{equation}
Here $\bar{y}^{\epsilon, x}(t)=\bar{y}^{\epsilon, x}(\theta_t\omega)$\,.
 Then we prove the following result

\begin{theorem}\label{thm:xe-x}
Assume~$\mathbf{H}_1$ and~$\mathbf{H}_2$. Given any $T>0$\,,
for any $x_0\in\R^n$, as $\epsilon\to0$ the solution of~(\ref{e:s-x}) converges in
probability in~$C(0, T; \R^n)$ to~$x$ which solves~(\ref{e:averaged}).
Moreover,  the rate of convergence is~$1/2$, that is,
\begin{equation*}
\sup_{0\leq t\leq T}\mathbb{E}|x^\epsilon(t)-x(t)|_{\R^n}\leq C_T\sqrt{\epsilon}
\end{equation*}
for some positive constant~$C_T$.
\end{theorem}

\begin{proof}
For any $T>0$\,, we partition~$[0, T]$ into subintervals of length
$\delta=\sqrt{\epsilon}$. Introduce a new process~$\tilde{x}^\epsilon(t)$
satisfying for $t\in[k\delta, (k+1)\delta)$\,, $k=0, 1, \ldots\,,
[T/\delta]$,
\begin{equation*}
\tilde{x}^\epsilon(t)=e^{(t-k\delta)A}x^\epsilon(k\delta)+\int_{k\delta}^te^{(t-s)A}f\left[x^\epsilon(k\delta),
\bar{y}^{\epsilon, x^{\epsilon}(k\delta)}(s)\right]ds\,.
\end{equation*}
Then for any $t\in[k\delta, (k+1)\delta)$
\begin{eqnarray*}
&&\mathbb{E}|x^\epsilon(t)-\tilde{x}^\epsilon(t)|_{\R^n}\\&\leq& \mathbb{E}\int_{k\delta}^t
e^{(t-s)A}\left|f\left[x^\epsilon(k\delta), \bar{y}^{\epsilon, x^{\epsilon}(k\delta)}(s)\right]-f\left[
x^\epsilon(s), \bar{y}^{\epsilon, x^\epsilon(s)}(s)\right]\right|_{\R^n}\,ds\\
&\leq& \mathbb{E}\int_{k\delta}^t
e^{(t-s)A}L_f\big[|x^\epsilon(k\delta)-x^\epsilon(s)|_{\R^n}+|\bar{y}^{\epsilon,
x^{\epsilon}(k\delta)}(s)-\bar{y}^{\epsilon,
x^\epsilon(s)}(s)|_{\R^m}\big]\,ds \\ &=&\mathbb{E}\int_{k\delta}^t
e^{(t-s)A}L_f\big[|x^\epsilon(k\delta)-x^\epsilon(s)|_{\R^n}+|h^\epsilon(x^{\epsilon}(k\delta),\theta_s\omega)-
h^\epsilon(x^\epsilon(s),\theta_s\omega)|_{\R^m}\big]\,ds
\end{eqnarray*}
\begin{eqnarray*}
&\leq&L_f(1+L_h)\int_{k\delta}^t
e^{(t-s)A}\mathbb{E}|x^\epsilon(k\delta)-x^\epsilon(s)|_{\R^n}\,ds\\
&\leq& C_T\delta
\end{eqnarray*}
for some positive constant~$C_T$.

On the other hand by~(\ref{e:y-ybar}) and noticing the Lipschitz property
of~$\bar{f}$ we have
\begin{eqnarray*}
&&\mathbb{E}|\tilde{x}^\epsilon(t)-x(t)|_{\R^n}\\
&\leq&\mathbb{E} \int_0^te^{(t-s)A}\left|f\left[x^\epsilon(\lfloor s/\delta \rfloor),
\bar{y}^{\epsilon, x^\epsilon(\lfloor s/\delta \rfloor)}(s)\right]-\bar{f}\left[x(s)\right]\right|_{\R^n}\,ds\\
&\leq & \mathbb{E}\int_0^te^{(t-s)A}\left|f\left[x^\epsilon(\lfloor s/\delta \rfloor),
\bar{y}^{\epsilon, x^\epsilon(\lfloor s/\delta \rfloor)}(s)\right]-
\bar{f}\left[x^\epsilon(\lfloor s/\delta \rfloor)\right]\right|_{\R^n}\,ds\\
&&{}+ \mathbb{E}\int_0^te^{(t-s)A}\left|\bar{f}\left[x^\epsilon(\lfloor s/\delta \rfloor)\right]-
\bar{f}\left[x^\epsilon(s)\right]\right|_{\R^n}\,ds\\
&&{}+\mathbb{E}\int_0^te^{(t-s)A}\left|\bar{f}\left[x^\epsilon(s)\right]-\bar{f}\left[x(s)\right]\right|_{\R^n}\,ds\\
&\leq & C_T\left[\sqrt{\epsilon}+\int_0^t\mathbb{E}|x^\epsilon(s)-x(s)|_{\R^n}\,ds\right]
\end{eqnarray*}
for some positive constant~$C_T$. Then we have
\begin{equation*}
\mathbb{E}|x^\epsilon(t)-x(t)|_{\R^n}\leq \mathbb{E}|x^\epsilon(t)-\tilde{x}^\epsilon(t)|_{\R^n}
+\mathbb{E}|\tilde{x}^\epsilon(t)-x(t)|_{\R^n}\,.
\end{equation*}
And by the Gronwall lemma we have for any $t\in[0, T]$,
\begin{equation*}
\mathbb{E}|x^\epsilon(t)-x(t)|_{\R^n}\leq C_T\sqrt{\epsilon}\,.
\end{equation*}
This completes the proof.
\end{proof}
Although the averaged equation is simple---being a deterministic autonomous system---the above result shows the approximation error is of~$\mathcal{O}(\sqrt{\epsilon})$.
Now in the following we give a refined approximation of~$x^\epsilon$
to errors of~$\mathcal{O}(\epsilon)$ for small $\epsilon>0$\,. For this define
\begin{equation*}
 H(x,t/\epsilon)=f\left(x, \bar{y}^{\epsilon, x}(t)\right)-\bar{f}(x).
\end{equation*}
Then rewrite~(\ref{e:reduced}) as
\begin{eqnarray}
\dot{x}^\epsilon(t)=Ax^\epsilon(t)+H(x^\epsilon(t), t/\epsilon)+\bar{f}(x^\epsilon(t)).
\end{eqnarray}
We follow a martingale approach to prove that $\int_0^t H(x^\epsilon(s), s/\epsilon)\,ds$
can be approximated by a Gaussian process. Define
\begin{equation}\label{e:bar-F}
\bar{H}(x, t)=\int_0^\infty \mathbb{E}\left[H(x, t+s)\given\mathcal{F}_t\right]\,ds
=\int_t^\infty\mathbb{E}\left[H(x, s)\given\mathcal{F}_t\right]\,ds
\end{equation}
and
\begin{eqnarray}\label{e:M-e}
M_t^\epsilon&=&\sqrt{\epsilon}\left[\bar{H}(x^\epsilon(t), t/\epsilon)-\bar{H}(x_0, 0)\right]
+\frac{1}{\sqrt{\epsilon}}\int_0^tH(x^\epsilon(s), s/\epsilon)\,ds
\\&&{}\quad
-\sqrt{\epsilon}\int_0^t\bar{H}_x(x^\epsilon(s), s/\epsilon)
\left[Ax^\epsilon(s)+H(x^\epsilon(s), s/\epsilon)+\bar{f}(x^\epsilon(s))\right]^*\,ds  \nonumber
\end{eqnarray}
where $*$~denotes the transpose of a vector.
Then we have the following result.
\begin{lemma}\label{lem:Me}
For each $\epsilon>0$\,, $\{M_t^\epsilon: 0\leq t\leq T\}$ is a martingale with respect to
$\{\mathcal{F}_{t/\epsilon}: t\geq 0\}$,  and the quadratic
covariance is
\begin{align}\label{e:covariance}
&\langle M^\epsilon \rangle_t = \epsilon\left[\bar{H}(x^\epsilon(t), t/\epsilon)\right]^2-
\epsilon\left[\bar{H}(x_0, 0)\right]^2-2\sqrt{\epsilon}\int_0^t\bar{H}(x^\epsilon(s), s/\epsilon)\,dM^\epsilon_s
\nonumber\\&{}
-2\epsilon\int_0^t\bar{H}(x^\epsilon(s), s/\epsilon)\bar{H}_x(x^\epsilon(s), s/\epsilon)\big[Ax^\epsilon(s)+H(x^\epsilon(s), s/\epsilon)+
\bar{f}(x^\epsilon(s))\big]^*\,ds
\nonumber\\&{}
+2\int_0^t\bar{H}(x^\epsilon(s), s/\epsilon)H^*(x^\epsilon(s), s/\epsilon)\,ds\,.
\end{align}
\end{lemma}

\begin{proof}
Let $[t_1, t_2]$ be an arbitrary subinterval of~$[0, T]$,
$\Delta=\{s_0, s_1, \ldots, s_N \}$ a partition of interval~$[t_1, t_2]$
 and $h=\max_{1\leq i\leq N}\{s_i-s_{i-1}\}$\,. Then
\begin{eqnarray*}
&&\mathbb{E}\left\{\bar{H}(x^\epsilon(t_2), t_2/\epsilon)-\bar{H}(x^\epsilon(t_1), t_1/\epsilon)
+\frac{1}{\sqrt{\epsilon}}\int_{t_1}^{t_2}H(x^\epsilon(s), s/\epsilon)\,ds
\right.\\&&\left.{}\quad-
\int^{t_2}_{t_1}\bar{H}_x(x^\epsilon(s), s/\epsilon)\left[Ax^\epsilon(s)
+H(x^\epsilon(s), s/\epsilon)+\bar{f}(x^\epsilon(s))\right]ds \Given
\mathcal{F}_{{t_1}/{\epsilon}} \right\}
\\
&=&\mathbb{E}\left\{\sum_{i=1}^N\int^{s_i}_{s_{i-1}}
\left[\bar{H}_x(x^\epsilon(s), s_i/\epsilon)-\bar{H}_x(x^\epsilon(s), s/\epsilon)\right]
\right.\\&&\left.{}\quad
 \times\left[Ax^\epsilon(s)
+H(x^\epsilon(s), s/\epsilon)+\bar{f}(x^\epsilon(s))\right] \Given\mathcal{F}_{\frac{t_1}{\epsilon}} \right\}
\\&&{}\quad
-\frac{1}{\sqrt{\epsilon}}\mathbb{E}\left\{\sum_{i=1}^N\int_{s_{i-1}}^{s_i}
\left[H(x^\epsilon(s_{i-1}), s/\epsilon)-H(x^\epsilon(s), s/\epsilon)\right]\,ds
\Given \mathcal{F}_{\frac{t_1}{\epsilon}} \right\}.
\end{eqnarray*}
By the assumption on~$f$ and the definition of~$\bar{H}$ we have
\begin{equation*}
\mathbb{E}\left\{\left|\bar{H}_x(x^\epsilon(s), s/\epsilon)-\bar{H}_x(x^\epsilon(s), s_i/\epsilon)\right|_{\R^n}\Given
\mathcal{F}_{t_1}\right\}\leq L_f h\,, \quad s\in [s_{i-1}, s_i],
\end{equation*}
and
\begin{eqnarray*}
&&\left|H(x^\epsilon(s_{i-1}), s/\epsilon)-H(x^\epsilon(s), s/\epsilon)\right|_{\R^n}\\
&\leq&\left|f(x^\epsilon(s_{i-1}), \bar{y}^{\epsilon, x^\epsilon(s_{i-1})}(s))-
f(x^\epsilon(s_{i-1}), \bar{y}^{\epsilon, x^\epsilon(s)}(s))\right|_{\R^n}
\\&&{}
+ \left |f(x^\epsilon(s_{i-1}), \bar{y}^{\epsilon, x^\epsilon(s)}(s))-
f(x^\epsilon(s), \bar{y}^{\epsilon, x^\epsilon(s)}(s)) \right |_{\R^n}
\\&&{}
+ \left |\bar{f}(x^\epsilon(s))-\bar{f}(x^\epsilon(s_{i-1})) \right |_{\R^n}
\\
&\leq& 2L_f(1+L_h) \left |x^\epsilon(s_{i-1})-x^\epsilon(s) \right |_{\R^n}\,.
\end{eqnarray*}
Then by the estimate in Lemma~\ref{lem:prior-est} and the Jensen
inequality for conditional expectation, passing to the limit $h\rightarrow 0$ yields
the first statement of the lemma.

By the It\^o formula
\begin{equation*}
d[M_t^\epsilon]^2=2M_t^\epsilon dM_t^\epsilon+d\langle M^\epsilon\rangle_t\,.
\end{equation*}
Then
\begin{equation*}
 \langle M^\epsilon\rangle_t=[M_t^\epsilon]^2-2\int_0^tM_s^\epsilon\, dM_s^\epsilon\,.
\end{equation*}
Substituting the expression for~$M^t_\epsilon$ into the above equation
and integrating by parts yields~(\ref{e:covariance}).
The proof is complete.
\end{proof}

Now by the definition of~$M_t^\epsilon$\,, we further rewrite~(\ref{e:reduced}) as
\begin{equation}\label{e:xe-Me}
x^\epsilon(t)=x_0+\int_0^t\left[Ax^\epsilon(s)+\bar{f}(x^\epsilon(s))\right]ds+\sqrt{\epsilon}M_t^\epsilon+\epsilon R^\epsilon(t),
\end{equation}
where
\begin{eqnarray*}
R^\epsilon(t)&=&-\bar{H}(x^\epsilon(t), t/\epsilon)+\bar{H}(x_0, 0)
\\&&{}+\int_0^t\bar{H}_x(x^\epsilon(s), s/\epsilon)\left[Ax^\epsilon(s)+H(x^\epsilon(s), s/\epsilon)+\bar{f}(x^\epsilon(s))\right]^*ds\,.
\end{eqnarray*}
By the estimate of Lemma~\ref{lem:prior-est}, the assumption on~$f$, and
the definition of~$\bar{H}$,
\begin{equation}\label{e:Re}
\lim_{\epsilon\rightarrow 0}\mathbb{E}\left[\sup_{0\leq t\leq T} \sqrt{\epsilon} R^\epsilon(t)\right]=0\,.
\end{equation}
Now define the process
\begin{equation}\label{e:M-M-e}
\mathcal{M}^\epsilon_t=\frac{1}{\sqrt{\epsilon}}\left\{x^\epsilon(t)-x_0-
\int_0^t\left[Ax^\epsilon(s)+\bar{f}(x^\epsilon(s))\right]ds \right\}.
\end{equation}
By the exponential mixing property of~$\bar{y}^{\epsilon, x}(t)$,
the assumption on~$f$, Lemma~\ref{lem:prior-est}
and~(\ref{e:xe-Me}), $\{\mathcal{M}^\epsilon_t\}_{0\leq \epsilon\leq
1}$ is tight in space~$C(0, T; \R^n)$. Let $P$~be a limit point of
the family of probability measures of
$\{\mathcal{L}(\mathcal{M}^\epsilon_t)\}_{0\leq \epsilon\leq 1}$\,.
And denote by~$\mathcal{M}^0_t$ a $C(0, T; \R^n)$-valued random
variable with distribution~$P$. Let $\Psi$~be a continuous bounded
function on~$C(0, T;\R^n)$. Set
$\Psi^\epsilon(s)=\Psi(x^\epsilon(s))$, then by~(\ref{e:xe-Me})
and~(\ref{e:Re}) and Lemma~\ref{lem:Me}
\begin{equation*}
 \mathbb{E}\left[\left(\mathcal{M}^\epsilon_t)-
\mathcal{M}^\epsilon_s)\right)\Psi^\epsilon(s)\right]
=\mathbb{E}\left[\left(\sqrt{\epsilon}R^\epsilon(t)-
\sqrt{\epsilon}R^\epsilon(s)\right)\Psi^\epsilon(s)\right]\rightarrow 0\,, \quad \epsilon\rightarrow 0\,,
\end{equation*}
which yields that the process~$\mathcal{M}_t^0$ is a $P$-martingale with respect
to the Borel $\sigma$-filter of~$C(0, T; \R^n)$. Further by Lemma~\ref{lem:Me}, we have
\begin{equation*}
 \mathbb{E}\left[\sup_{0\leq t\leq T}\Big|\langle M^\epsilon\rangle_t-
2\int_0^t\bar{H}(x^\epsilon(s), s/\epsilon)H^*(x^\epsilon(s), s/\epsilon)\,ds\Big|\right]
\rightarrow 0\,, \quad \epsilon\rightarrow 0\,.
\end{equation*}
Denote by~$\Sigma(x)$ the matrix $2\mathbb{E}\left[\bar{H}(x, t)H^*(x, t)\right]$, $x\in\R^n$,
which is independent of~$t$. By the definition of~$H$, $\bar{H}$ and
the exponential mixing property of~$\bar{y}^{\epsilon, x}$, we derive the
following representation for~$\Sigma(x)$:
\begin{eqnarray}
\Sigma(x)&=&2\mathbb{E}\left[\int_t^\infty \mathbb{E}[H(x, s)\given\mathcal{F}_t]\,ds \, H^*(x, t)\right]\nonumber\\
&=&2\int_0^\infty\mathbb{E}\left[H(x, s+t)H^*(x, t)\right]ds\nonumber\\
&=&2\int_0^\infty\mathbb{E}\left[H(x, s)H^*(x, 0)\right]ds\,. \label{Sigma}
\end{eqnarray}
Notice that~$\bar{H}(\cdot, t)$ and~$H(\cdot, t)$ are Lipschitz
uniformly with respect to~$t$. So we follow the approach used
by Pardoux et al.~\cite[Proposition 8]{PP03}:
\begin{equation*}
\mathbb{E}\left\{\sup_{0\leq t\leq
T}\left|\int_0^t\left[\bar{H}(x^\epsilon(s),
s/\epsilon)H^*(x^\epsilon(s), s/\epsilon)- \Sigma(x^\epsilon(s))
\right]ds \right|  \right\}\rightarrow 0\,, \quad
\epsilon\rightarrow 0\,.
\end{equation*}
Then we pass limit $\epsilon\rightarrow 0$ in~(\ref{e:M-M-e}) and arrive at the following
result.
\begin{lemma}
The process
\begin{equation*}
\mathcal{M}^0_t=\lim_{\epsilon\rightarrow 0}\mathcal{M}_t^\epsilon=\lim_{\epsilon\rightarrow 0}
\frac{1}{\sqrt{\epsilon}}\left\{x^\epsilon(t)-x_0-
\int_0^t\left[Ax^\epsilon(s)+\bar{f}(x^\epsilon(s))\right]ds \right\}
\end{equation*}
defined on the probability space $(C(0, T; \R^n),
\mathcal{B}(C(0, T; \R^n)), P)$ is  a square integrable martingale
with the associated quadratic variation process given by~$\Sigma(x(t))$ which is defined by~(\ref{Sigma}).
\end{lemma}
\begin{proof}
It is just necessary to prove the uniqueness of the martingale problem.
By the martingale representation theorem \cite{IW81},
without changing the distribution of~$\mathcal{M}_t^\epsilon$ and~$\mathcal{M}_t^0$,  one can enlarge the original probability space to
($\tilde{\Omega}$, $\tilde{\mathcal{F}}$, $\mathcal{F}_t$,  $\tilde{\mathbb{P}}$)
and there is a $\R^n$-valued Wiener process~$\tilde{W}(t)$ such
that $\mathcal{M}_t^0$~satisfies
\begin{equation*}
 dz(t)=\bar{\sigma}(x(t)) d\tilde{W}(t), \quad z(0)=0 \,,
\end{equation*}
where $\bar{\sigma}(x)\in\R^{n\times n}$ such that
$\bar{\sigma}(x)\bar{\sigma}^*(x)=\Sigma(x)$, $x(t)$
solves~(\ref{e:averaged}). By the definition of~$\Sigma$ and~$H$,
$\sigma$~is Lipschitz which yields the existence and uniqueness of
$P$\,. This completes the proof.
\end{proof}

Now in order to derive the intermediate reduced system we introduce~$z^\epsilon$ which
solves
\begin{equation*}
dz^\epsilon(t)=\bar{\sigma}(x^\epsilon(t))d\tilde{W}(t), \quad
z^\epsilon(0)=0\,.
\end{equation*}
Then by the Burkholder--Davis--Gundy inequality~\cite{Huang} and
 Theorem~\ref{thm:xe-x}, we have for any $T>0$ that
\begin{equation*}
\tilde{\mathbb{E}}\sup_{0\leq t\leq T}|z^\epsilon(t)-z(t)|^2_{\R^n}\leq L_{\sigma}C_T\,\epsilon
\end{equation*}
for some positive constant~$C_T$. Here $\tilde{\mathbb{E}}$~is the expectation operator defined on the enlarged probability
space with respect to probability measure~$\tilde{\mathbb{P}}$. Then by the definition of~$\mathcal{M}^0_t$, without changing
the distribution of~$x^\epsilon$, over the enlarged probability space
$(\tilde{\Omega}, \tilde{\mathcal{F}}, \tilde{\mathbb{P}})$
we have
\begin{equation*}
x^\epsilon(t)=x_0+\int_0^t\left[Ax^\epsilon(s)+\bar{f}(x^\epsilon(s))\right]ds+
\sqrt{\epsilon}\int_0^t\bar{\sigma}(x^\epsilon(s))\,d\tilde{W}(s)+\mathcal{O}(\epsilon).
\end{equation*}
By discarding  terms of~$\mathcal{O}(\epsilon)$ we derive the
following reduced approximation system
\begin{equation*}
\tilde{x}^\epsilon(t)=x_0+\int_0^t\left[A\tilde{x}^\epsilon(s)+\bar{f}(\tilde{x}^\epsilon(s))\right]ds+
\sqrt{\epsilon}\int_0^t\bar{\sigma}(\tilde{x}^\epsilon(s))\,d\tilde{W}(s),
\end{equation*}
or, equivalently, the following differential form
\begin{equation}\label{e:inter-reduced}
d\tilde{x}^\epsilon(t)=\left[A\tilde{x}^\epsilon(t)+\bar{f}(\tilde{x}^\epsilon(t))\right]dt
+\sqrt{\epsilon}\bar{\sigma}(\tilde{x}^\epsilon(t))\,d\tilde{W}(t),
\quad \tilde{x}^\epsilon(0)=x_0\,.
\end{equation}
Then we draw the following result
\begin{theorem}
Assume~$\mathbf{H}_1$ and~$\mathbf{H}_2$. Over the enlarged
probability space $(\tilde{\Omega}, \tilde{\mathcal{F}},
\tilde{\mathbb{P}})$, without changing the distribution, for any
$T>0$\,, $x_0\in\R^n$, the solution of~(\ref{e:s-x}), $x^\epsilon$,
is approximated by~$\tilde{x}^\epsilon$, the solution
of~(\ref{e:inter-reduced}), in the space~$C(0, T; \R^n )$ up to
errors of~$\mathcal{O}(\epsilon)$, that is almost surely
\begin{equation*}
|x^\epsilon-\tilde{x}^\epsilon|_{C(0, T; \R^n)}=\mathcal{O}(\epsilon), \quad \epsilon\rightarrow 0\,.
\end{equation*}
\end{theorem}

\begin{remark}
By the above result, the system~(\ref{e:inter-reduced})---which is a
deterministic equation with small stochastic perturbation---is an
intermediate reduced system for the original slow-fast system~(\ref{e:s-x})--(\ref{e:s-y}). Moreover, the system~(\ref{e:inter-reduced}) approximates original system~(\ref{e:s-x})--(\ref{e:s-y}) up to errors of~$\mathcal{O}(\epsilon)$.
\end{remark}

For an example, we now return to the simple model of Section~\ref{sec:toy model}. By
the definition of~$f$ and~$g$ in Section~\ref{sec:toy model}, for
any fixed $x\in\R$\,, (\ref{e:toy-fast2})~has a unique stationary
solution~$\bar{y}^{\epsilon,x}$ which is exponential mixing. Then
near $x=0$ we have the asymptotic expansion
\begin{equation*}
\bar y^{\epsilon, x}=x^2-\sigma^2+\mathcal{O}(x^3, \sqrt{\epsilon}).
\end{equation*}
And the averaged  equation near $x=0$ has the following asymptotic
form
\begin{equation}
\dot{x}=\bar{f}(x)=-x^3+\sigma^2x+\mathcal{O}(x^4,\sqrt{\epsilon}).
\end{equation}
To obtain an intermediate reduced system we calculate, near $x=0$\,,
\begin{eqnarray*}
&&2\mathbb{E}\int_0^\infty\big[-x\bar y^{\epsilon, x}(t)+x\mathbb{E}\bar
y^{\epsilon,x}(t)\big]\big[-x\bar y^{\epsilon, x}(0)+x\mathbb{E}\bar
y^{\epsilon,x}(0)\big]\,dt\\
&=&2x^2\int_0^\infty\big[\mathbb{E}\big(\bar y^{\epsilon,x}(t)\bar
y^{\epsilon, x}(0)\big)-\mathbb{E}\big(\bar y^{\epsilon,x}\big)^2
\big]\,dt\\ &=&
x^2\sigma^2-12\sigma^2x^4+20\sigma^4x^2+\mathcal{O}(x^3).
\end{eqnarray*}
Then we obtain the following reduced system
\begin{equation*}
dx=[-x^3+\sigma^2x]\,dt+ \sigma
x\sqrt{\epsilon}\sqrt{1-12x^2+20\sigma^2}\,dW(t).
\end{equation*}

\section{Conclusion}

Averaging of systems with noise in only the fast variables leads to a deterministic reduced model with errors~$\mathcal O(\sqrt\epsilon)$, Section~\ref{sec:average}.  The stochastic slow manifold, Section~\ref{sec:reduction on IM}, shows that generally there should be noise in the reduced model, as seen in the simple example of Section~\ref{sec:toy model}.  Consequently, Section~\ref{sec:average}, shows that a martingale argument establishes the leading influence of the fast noise on the averaged system.  When added to the averaged system the resultant reduced model typically has the smaller error~$\mathcal O(\epsilon)$.  This work not only shows how to improve averaged models of stochastic systems, but strengthens the connections between the methodologies of averaging and slow manifold reduction.

\paragraph{Acknowledgements} This research is supported by the
Australian Research Council grant DP0774311 and NSFC grant 10701072.

\end{document}